\documentclass[letterpaper]{article} 
\usepackage{aaai25}  
\usepackage{times}  
\usepackage{helvet}  
\usepackage{courier}  
\usepackage[hyphens]{url}  
\usepackage{graphicx} 
\usepackage{natbib}  
\usepackage{caption} 
\usepackage{algorithm}
\usepackage{algorithmic}

\usepackage{newfloat}
\usepackage{listings}
\DeclareCaptionStyle{ruled}{labelfont=normalfont,labelsep=colon,strut=off} 
\floatstyle{ruled}
\newfloat{listing}{tb}{lst}{}
\floatname{listing}{Listing}
\usepackage{booktabs}
\usepackage{amssymb}

\title{KD-MSLRT: Lightweight Sign Language Recognition Model Based on Mediapipe and 3D to 1D Knowledge Distillation}

\author{
	Yulong Li\textsuperscript{\rm 1}\equalcontrib,
	Bolin Ren\textsuperscript{\rm 2}\equalcontrib,
	Ke Hu\textsuperscript{\rm 3},
	Changyuan Liu\textsuperscript{\rm 4},
	Zhengyong Jiang\textsuperscript{\rm 5},
	Kang Dang\textsuperscript{\rm 6}\thanks{Corresponding authors.},
	Jionglong Su\textsuperscript{\rm 7}\textsuperscript{\dag}
}

\affiliations{
	\textsuperscript{\rm 1,2,4,5,6,7}School of AI and Advanced Computing, XJTLU Entrepreneur College (Taicang), Xi'an Jiaotong-Liverpool University, Suzhou, 215123, China\\
	\textsuperscript{\rm 3}School of Humanities and Social Sciences, Xi’an Jiaotong-Liverpool University, Suzhou, 215123, China\\
	\{\textsuperscript{}Kang.Dang, \textsuperscript{}Jionglong.Su\}@xjtlu.edu.cn
}

\begin{document}

\maketitle

\begin{abstract}
	Artificial intelligence has achieved notable results in sign language recognition and translation. However, relatively few efforts have been made to significantly improve the quality of life for the 72 million hearing-impaired people worldwide. Sign language translation models, relying on video inputs, involves with large parameter sizes, making it time-consuming and computationally intensive to be deployed. This directly contributes to the scarcity of human-centered technology in this field. Additionally, the lack of datasets in sign language translation hampers research progress in this area. To address these, we first propose a cross-modal multi-knowledge distillation technique from 3D to 1D and a novel end-to-end pre-training text correction framework. Compared to other pre-trained models, our framework achieves significant advancements in correcting text output errors. Our model achieves a decrease in Word Error Rate (WER) of at least 1.4\% on PHOENIX14 and PHOENIX14T datasets compared to the state-of-the-art CorrNet. Additionally, the TensorFlow Lite (TFLite) quantized model size is reduced to 12.93 MB, making it the smallest, fastest, and most accurate model to date. We have also collected and released extensive Chinese sign language datasets, and developed a specialized training vocabulary. To address the lack of research on data augmentation for landmark data, we have designed comparative experiments on various augmentation methods. Moreover, we performed a simulated deployment and prediction of our model on Intel platform CPUs and assessed the feasibility of deploying the model on other platforms.
\end{abstract}

%
\renewcommand{\thefootnote}{}

\section{Introduction}
72 million hearing-impaired individuals worldwide rely on sign language as their primary mode of communication, as estimated by the World Federation of the Deaf~\cite{1}.  However, the lack of widespread sign language understanding creates significant barriers to their full participation in society. While the rapid advancements in Natural Language Processing (NLP) and Computer Vision (CV) have made video-based sign language recognition (SLR) more feasible~\cite{2}, existing sign language recognition (SLR) models often fall short in addressing the practical needs of the hearing-impaired community.

\begin{figure}[t]
	\centering
	\includegraphics[width=1.1\columnwidth]{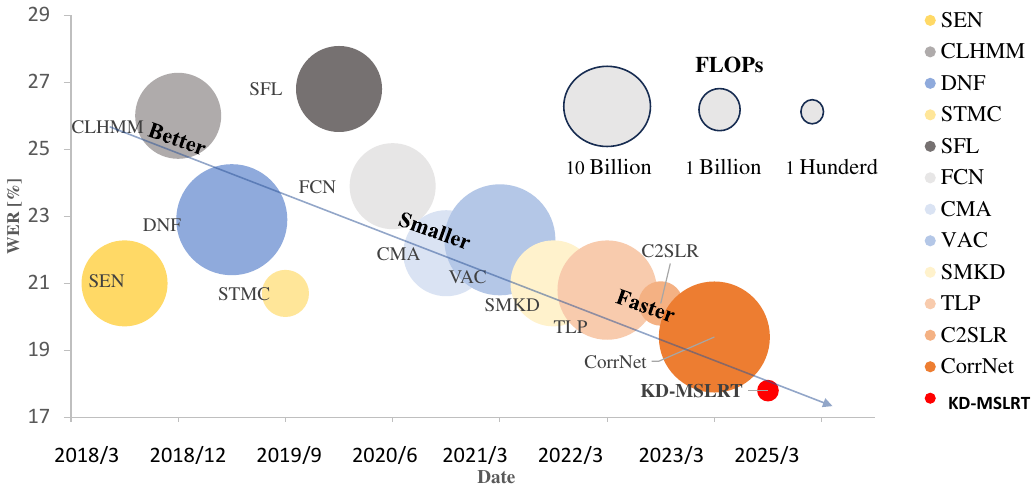}
	\caption{Publication time versus model performance. Larger circles indicate higher FLOPs, reflecting greater computational complexity.}
	\label{fig1}
    \vspace{-2em}
\end{figure}

Current SLR approaches, primarily based on complex video analysis, face three major limitations.  First, these models are computationally expensive~\cite{7}, hindering their deployment on resource-constrained devices like mobile phones, crucial for real-time communication accessibility.  Second, despite significant progress in American Sign Language (ASL) recognition, research on other sign languages, particularly Chinese Sign Language (CSL), remains limited due to a lack of publicly available datasets ~\cite{5}. Third, even state-of-the-art models are also prone to output errors, requiring additional mechanisms for grammatical and contextual correction to ensure accurate and understandable translations.

Our research addresses these challenges by introducing KD-MSLRT, a novel \textbf{K}nowledge \textbf{D}istillation-Based \textbf{M}ediaPipe \textbf{S}ign \textbf{L}anguage \textbf{R}ecognition model with \textbf{T}ext Correction. Our approach prioritizes both accuracy and efficiency, aiming for practical deployment in real-world scenarios. Instead of relying on computationally intensive video processing~\cite{6}, KD-MSLRT leverages MediaPipe~\cite{8}, a powerful framework for real-time human pose estimation. By extracting skeletal landmarks from sign language videos, we simplify the input data from 3D to 1D, significantly reducing computational burden without compromising accuracy~\cite{9}. This allows for efficient processing on even resource-constrained devices~\cite{10, hu2025ophnet}, making real-time sign language translation accessible to a wider audience. Figure \ref{fig1} shows the publication time and performance of various models. Compared to other recently released baseline models, KD-MSLRT leads in both Word Error Rate (WER) and FLOPs. This can be attributed to three key innovations to enhance KD-MSLRT's performance:
\begin{itemize}
	\item \textbf{3D to 1D Knowledge Distillation}. This technique enables our lightweight model to learn from a more complex and accurate video-based model \cite{11}, achieving comparable performance with significantly reduced resource requirements.
	\item \textbf{Novel Temporal and Spatial Data Augmentation Algorithms}. Specifically designed for landmark-based sign language data, these algorithms significantly improve the model's ability to generalize and recognize diverse signing styles.
	\item \textbf{Specialized Text Correction Network}. A novel text correction model trained on a self-supervised approach ensures grammatically correct and contextually relevant translations, addressing the common issue of errors in language model outputs.
\end{itemize}
Recognizing the scarcity of CSL resources, we also introduce a new large-scale dataset of 8,976 samples, focusing on long and complex sentences commonly found in news broadcasts. This dataset will be a valuable resource for researchers and developers, fostering further advancements in CSL recognition.

Our contributions are as follows:
\begin{itemize}
	\item We developed KD-MSLRT, a highly accurate and efficient sign language recognition model that achieves a decrease in WER of at least 1.4\% on PHOENIX14 and PHOENIX14T compared to state-of-the-art (SOTA) models. With a TensorFlow Lite (TFLite) quantized model size of just 12.93 MB, it can be deployed by enterprises at minimal cost.
	\item To the best of our knowledge, we are the first to propose temporal and spatial data augmentation algorithms specifically for landmark-format sign language data, which significantly enhances the effectiveness of landmark-based sign language recognition models.
	\item We collect and release the largest Chinese sign language recognition dataset in the CSLR field, containing 8,976 Chinese sign language samples with corresponding labels of complete news sentences. 
	\item We performed a simulated deployment and prediction of our model on Intel platform CPUs and assessed the feasibility of deploying the model on other platforms.
\end{itemize}

\section{Related work}
\subsubsection{Continuous Sign Language Recognition.}
Continuous Sign Language Recognition (CSLR) is the recognition of consecutive sign language videos as an annotated sequence of sign languages after manual annotation \cite{19,20}.  Early methods in CSLR \cite{24, 25} always use manually crafted features or systems based on Hidden Markov Models \cite{26} to perform temporal modeling and gradually translate sentences. HMM-based systems first use feature extractors to capture visual features and then use HMMs to perform long-term temporal modeling \cite{27}. With the introduction of Connectionist Temporal Classification (CTC) loss \cite{25}, significant improvements were achieved in the field of speech recognition. It can also be applied in CSLR \cite{28}, i.e., by sequentially aligning the target sentence with the input frames. The basis for using CTC loss in CSLR lies in extracting semantic information from sign language videos with long-term temporal dependencies. This requires the feature extractor to have a subtle structural design and to ensure that the spatial information is accurately extracted from the spatial information to the spatial features that contain the semantic information \cite{wang2022stepwise,zhao2024sfc}. Therefore, Correlation Network is proposed \cite{13}, which introduces Correlation operation as a spatial feature extractor based on a ResNet network, and employs 1D CNN and bidirectional LSTM (BiLSTM) for short-term and long-term temporal modeling, respectively, and employs classifiers to predict sentences.

With the proposal of models such as MediaPipe for extracting pose keypoints, C2SLR \cite{29} uses pre-extracted pose keypoints as supervision to guide the model to focus explicitly on hand and face regions. However, as the model structure becomes more complex, the number of parameters and required computational resources has grown exponentially, leading to challenges in industrial applications. To address these, we propose a lightweight and fast sign language recognition model based on MediaPipe.

\begin{figure*}[t]
	\centering
	\includegraphics[width=1.8\columnwidth]{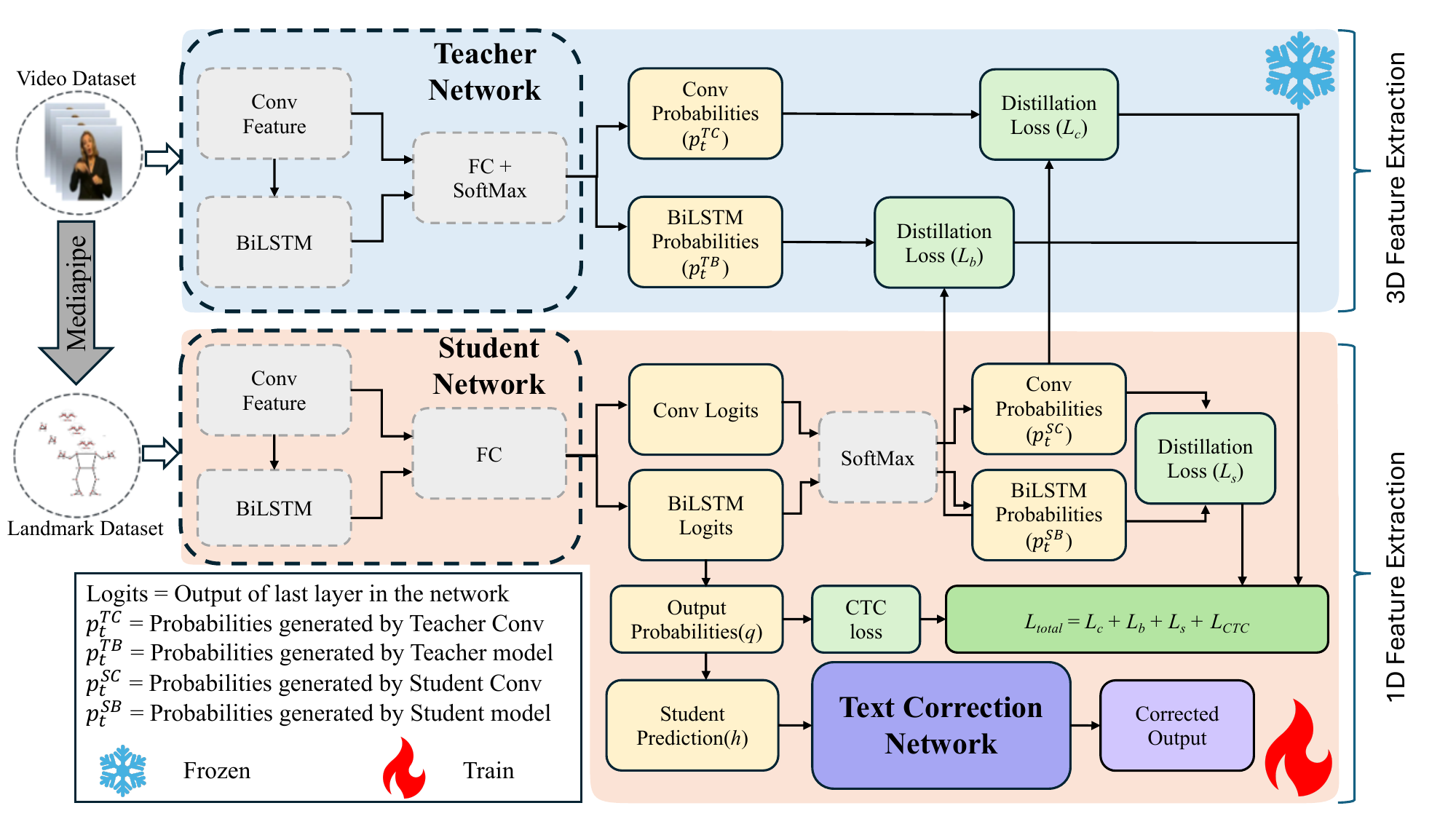} 
	\caption{An overview of our KD-MSLRT. We adopt the SOTA video-based sign language model CorrNet \cite{12} as the teacher network. We conduct knowledge distillation separately on the probabilities obtained from the convolutional layers and bidirectional LSTM. The probabilities from the bidirectional LSTM are used to predict sentences, followed by further refinement of the predicted sentences using a text correction network trained on a large corpus.}
	\label{fig2}
\end{figure*}

\subsubsection{Application of MediaPipe to CSLR.}
OpenPose \cite{30} and MediaPipe Holistic \cite{8} models, which are fast and highly accurate, have achieved remarkable results in the field of pose recognition. They can quickly and accurately extract the airspace information of key nodes such as hands, face, and pose in the video. Hence, OpenPose and MediaPipe are applied in ISLR modeling \cite{31,32}. Models using human body boundary markers as data require fewer parameters than video-based models, and the model architecture is simpler without the need for complex model structure design \cite{33}. The rapid development of the field of sign language recognition (SLR) and interpreting cannot be promoted without commercialized applications. Therefore, a SLR model based on MediaPipe is more suitable for commercial deployment, and MediaPipe supports fast deployment on mobile devices such as Android and iOS. 

However, SLR models based on landmarks are far inferior performance-wise to other models that are based on video \cite{34}. The primary reason for this performance gap is that landmark data only includes spatial information about hand and body movements, whereas video-based sign language often includes more contextual information, such as facial expressions, body posture, and temporal dynamics, which are crucial for accurate interpretation. Relying solely on the limited information provided by landmarks may not fully capture the entire linguistic and semantic context conveyed by sign language. To address this, we propose a 3D to 1D knowledge distillation technique and a self-supervised learning text correction model.

\subsubsection{Text correction network.}
Recent years have seen significant advancements in the research on text correction within the field of NLP \cite{TC1,TC2}. The introduction of sequence-to-sequence (Seq2Seq) models \cite{TC3} has led to substantial advancements in this field, particularly in the development of text correction networks. Seq2Seq-based approaches, with their ability to manage complex sequence alignment and transformation, have established a robust framework for addressing textual errors \cite{TC4}. Furthermore, Transformer-based models have significantly advanced the SOTA in text correction \cite{TC5}. BERT (Bidirectional Encoder Representations from Transformers) \cite{TC6} and its derivatives have been extensively utilized for pre-training text correction systems, leveraging their capacity to capture rich contextual semantics \cite{TC7}. However, models based on these two architectures can encounter issues when dealing with random errors, as a single error in the input sequence may lead to a cascade of errors in the output sequence. This error propagation effect is particularly pronounced when processing longer sentences or more complex structures. Additionally, these models primarily rely on patterns and rules learned during pre-training \cite{tang2024hunting, hu2024ophclip}. They may fail to recognize the nature of random errors, potentially misinterpreting them as signals. To address these challenges, we propose a novel end-to-end self-supervised pre-training network specifically designed to correct random errors in the output text of language translation and recognition models.

\section{Methodology}

This research introduces the \textbf{K}nowledge \textbf{D}istillation-Based \textbf{M}ediaPipe \textbf{S}ign \textbf{L}anguage \textbf{R}ecognition model with \textbf{T}ext Correction Network (KD-MSLRT), a lightweight and efficient sign language recognition model designed for resource-constrained environments. KD-MSLRT leverages a novel combination of 3D to 1D knowledge distillation, a MediaPipe-based architecture, landmark data augmentation, and a novel text correction network to address the challenges of accurate and efficient sign language recognition.

\subsection{Overall Network Design and Knowledge Distillation}

KD-MSLRT comprises three primary modules, as illustrated in Figure \ref{fig2}: a robust teacher network, a lightweight student network, and a novel text correction network.

The objective of both the teacher and student networks in this research is to translate the input video or extracted landmark data $X = \{ x_t \}_{t=1}^{T}$ into a series of glosses $Y = \{ y_i \}_{i=1}^{N}$ representing a sentence, where $N$ denotes the length of the glosses label, $T$ represents the time frame. Finally, the classifier uses CTC loss to predict the probability of the target gloss sequence $p(Y|X)$. 

The teacher network utilizes CorrNet, a state-of-the-art model for video-based Continuous Sign Language Recognition (CSLR), to provide robust knowledge for the student network. CorrNet operates on individual frames of a sign language video, effectively extracting features and capturing temporal dependencies without explicitly modeling frame continuity. Its key components include a Feature Extraction Layer, a 1D CNN for capturing spatial relationships within frames, a BiLSTM for modeling temporal dependencies between frames, and a Classification Layer for predicting the probability of a gloss sequence. Furthermore, CorrNet incorporates a Correlation Module for identifying body trajectories and an Identification Module for focusing on information-rich regions.

\begin{figure}[t]
	\centering
	\includegraphics[width=1\columnwidth]{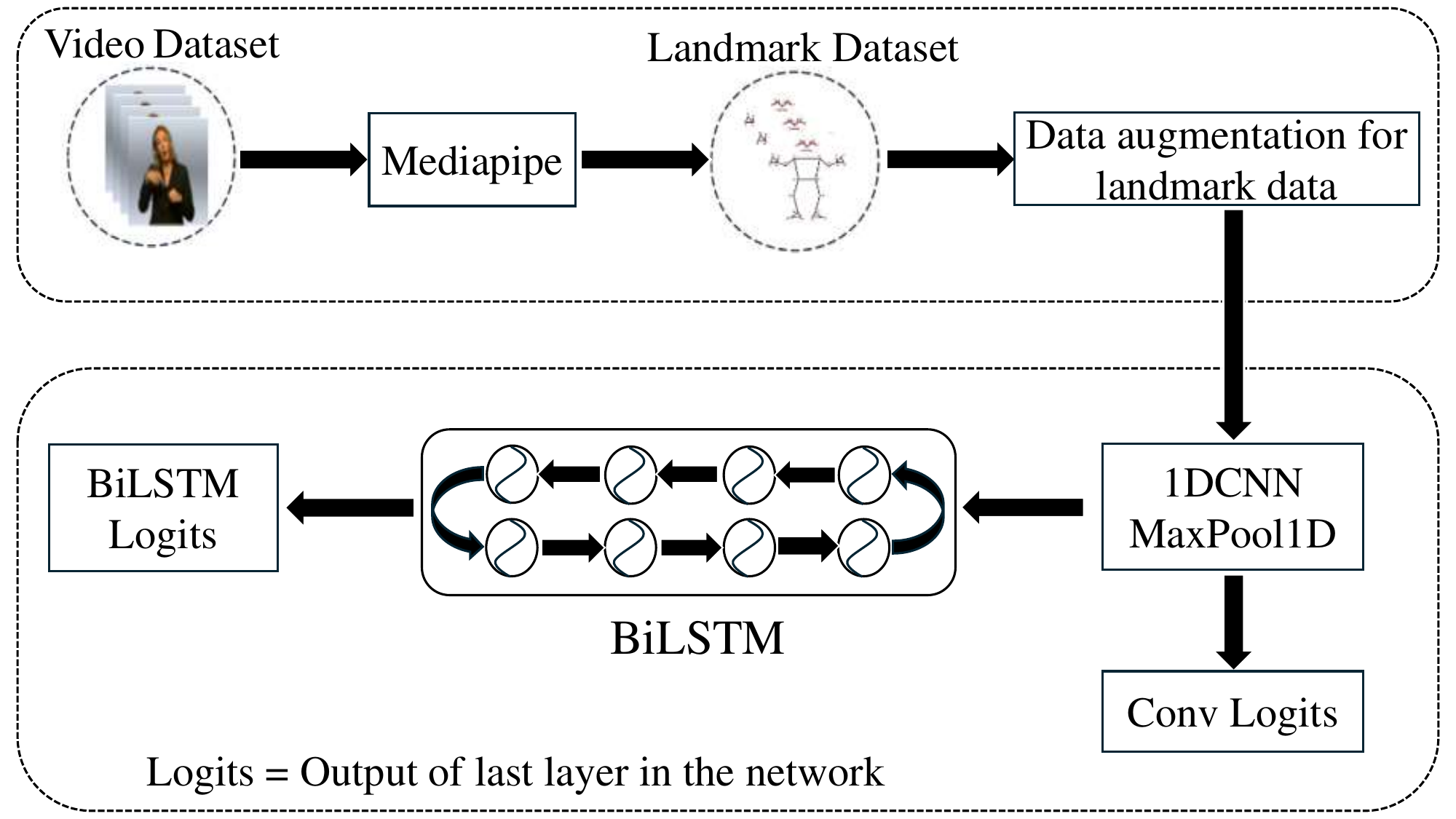} 
	\caption{The lightweight sign language recognition model MSLR proposed based on MediaPipe in this research.}
	\label{fig3}
    \vspace{-2.5em}
\end{figure}

The student network, named the \textbf{M}ediaPipe \textbf{S}ign \textbf{L}anguage \textbf{R}ecognition model (MSLR), is designed for lightweight and efficient sign language recognition. MSLR operates on landmark data extracted from videos using MediaPipe, significantly reducing the computational burden while preserving essential information for sign recognition. As shown in Figure \ref{fig3}, MSLR consists of a Convolutional Feature Extraction Layer (four layers of 1D CNN) and a BiLSTM Layer for capturing temporal dependencies, mirroring the teacher network's structure. Its input is the human landmark data extracted by MediaPipe. The original video data $X_{{video}} = \{ x_t \}_{t=1}^{T} \in {\mathbb{R}}^{T \times C \times H \times W}$ is processed by MediaPipe to obtain $X_{{landmark}} = \{ x_t \}_{t=1}^{T} \in {\mathbb{R}}^{T \times C \times F}$, where $F$ is the selected key nodes. In this research, only 72 keypoints, including hand and partial facial landmarks, are used as input to the student network, so ${F = 72}$. 

KD-MSLRT employs a 3D to 1D knowledge distillation strategy, transferring knowledge from the teacher network's convolutional and BiLSTM layers to the student network. This process encourages the student network to learn discriminative features and temporal dependencies similar to the teacher network, enhancing its recognition capabilities. Additionally, both networks share a final classification layer and undergo self-knowledge distillation, further refining the learned feature representations.

The knowledge distillation process is implemented by minimizing a combined loss function that considers the distillation losses from the convolutional layer, BiLSTM layer, and self-knowledge distillation, as defined in Equations 1-5.

\begin{equation}
	{\mathcal{L}}_{c} = \alpha \sum_{t=1}^{T} \left( p_t^{TC} \left( \log p_t^{TC} - p_t^{SC} \right) \right),
\end{equation}
\begin{equation}
	{\mathcal{L}}_{b} = \alpha \sum_{t=1}^{T} \left( p_t^{TB} \left( \log p_t^{TB} - p_t^{SB} \right) \right),
\end{equation}
\begin{equation}
	{\mathcal{L}}_{s} = \alpha \sum_{t=1}^{T} \left( p_t^{SB} \left( \log p_t^{SB} - p_t^{SC} \right) \right),
\end{equation}
\begin{equation}
	{\mathcal{L}}_{CTC}(X, Y) = -\log p(Y\mid X),
\end{equation}
\begin{equation}
	{\mathcal{L}}_{total} = {\mathcal{L}}_{c} + {\mathcal{L}}_{b} + {\mathcal{L}}_{s} + {\mathcal{L}}_{CTC},
\end{equation}
where ${\mathcal{L}}_{c},{\mathcal{L}}_{b},{\mathcal{L}}_{s}$ are the distillation loss of the convolutional layer, BiLSTM and self-knowledge distillation, respectively, $T$ is the time frame corresponding to the input, $\alpha$ is the temperature parameter, and $p_t^{TC}, p_t^{TB}, p_t^{SC}, p_t^{SB}$ are the probability outputs of different modules as defined in Figure \ref{fig2}.

\subsection{Landmark Data Augmentation}

To enhance the robustness and generalization ability of the student network, KD-MSLRT introduces landmark data augmentation techniques. These techniques augment the training data by introducing variations in landmark positions and sequences, simulating real-world scenarios and improving the network's ability to generalize to unseen data.

To enhance robustness and generalization, three landmark data augmentation techniques are employed: spatial rotation, which simulates variations in camera viewpoint and signer orientation by rotating landmark coordinates; random translation, which accounts for signer positioning variations within the video frame by shifting the entire landmark sequence; and frame fusion, which generates new frames by combining information from consecutive frames using either average fusion (averaging landmark coordinates) or weighted fusion (combining landmarks with varying degrees of influence).

These augmentation techniques, aligned with the video augmentation techniques used for the teacher network, ensure consistency in the training process and contribute to the student network's robustness and accuracy.

\subsection{Text Correction Network}

The final component of KD-MSLRT is the text correction network, designed to refine the output from the student network and address potential recognition errors. This network identifies and corrects common errors such as misidentifications, omissions, insertions, and disordered sequences in the recognized gloss sequences, ensuring a more accurate and grammatically correct final translation.

\begin{figure}[t]
	\centering
	\includegraphics[width=1\columnwidth]{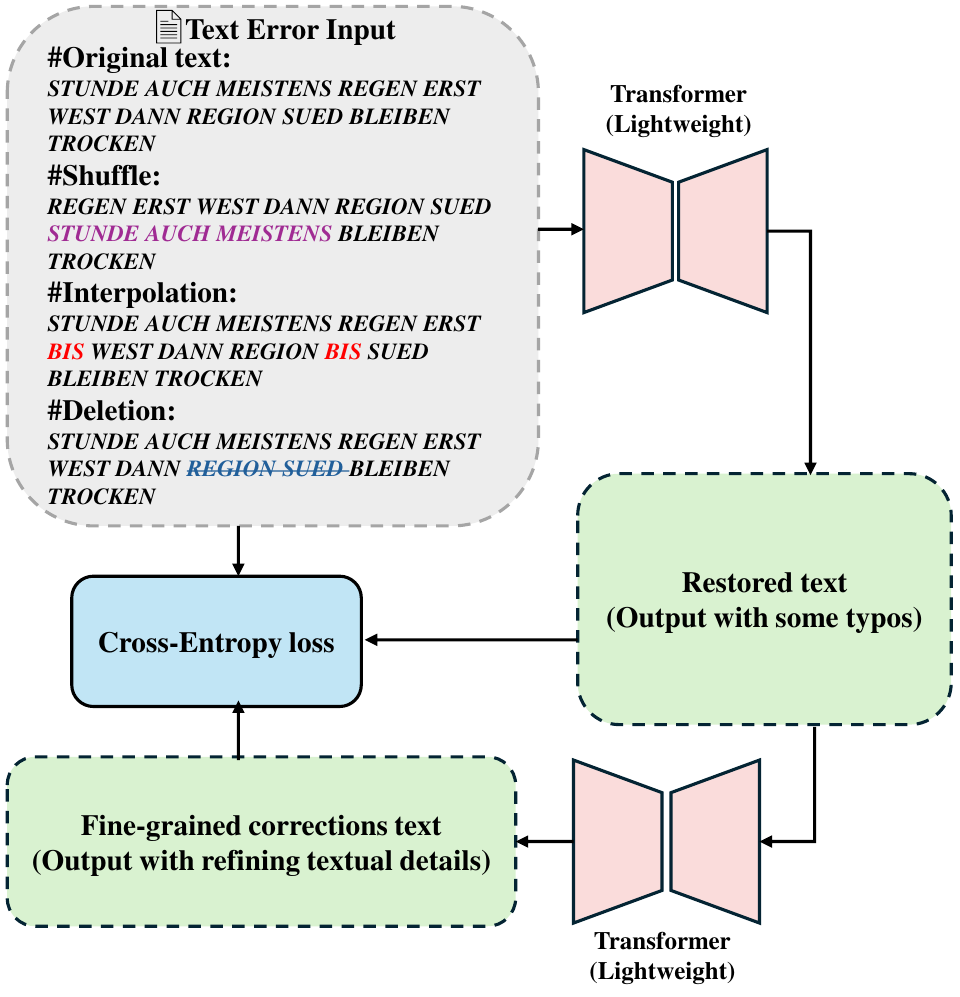}
	\caption{The self-supervised training text correction model proposed based on the error types of the SLR model outputs in this research.}
	\label{fig4}
\end{figure}

\begin{figure}[t]
	\centering
	\includegraphics[width=\columnwidth]{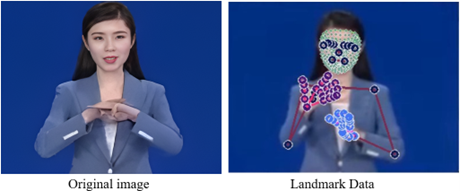}
	\caption{The Chinese long sentence sign language dataset collected for this release consists entirely of news content, including a rich array of professional terms.}
	\label{fig5}
\end{figure}

As shown in Figure \ref{fig4}, the text correction network consists of two lightweight Transformer models with a high-encoding, low-decoding structure, enabling efficient real-time performance on resource-constrained devices. The first Transformer Module focuses on restoring the original meaning of the recognized text, preserving the overall intent of the sentence, while the second Transformer Module performs fine-grained text corrections, addressing remaining errors.

A novel self-supervised pre-training approach is employed to train the text correction network effectively. This approach generates training data by introducing artificial errors into correct text, simulating the types of errors commonly made by the student network. This allows the network to learn to identify and rectify similar errors in the actual recognized text, significantly enhancing the accuracy and readability of the final sign language translations.

Our self-supervised pre-training process utilizes both original and noise-augmented text versions.  Noise augmentation includes techniques like shuffling, random deletion, and insertion of sequences.  This combined approach offers several benefits:
\begin{itemize}
	\item Preserves Meaning: The original text helps the model learn full word relationships and sentence meaning.
	\item Expands Training Data: Combining original and modified texts exposes the model to more word combinations, improving its ability to handle new sentences.
	\item Prevents Overfitting: Including original text ensures the model learns actual semantics, not just patterns from preprocessing.
	\item Enhances Word Understanding: Direct input of original text helps the model capture word usage in different contexts.
\end{itemize}

This approach ultimately enhances the model's understanding of sentence structure and its ability to interpret relationships between sentences.  It also increases the model's robustness, making it more resilient to noisy or incomplete input.  By learning to balance local word relationships with the overall sentence structure, the model can effectively handle complex outputs and correct errors in challenging scenarios, particularly when dealing with noisy data.

\subsection{Chinese Standard Sign Language Dataset}

This research addresses the scarcity of publicly available Chinese Standard Sign Language data, particularly for recognizing long sentences, by introducing a new dataset collected from the official Chinese TV channel program Common Concern, as illustrated in Figure \ref{fig5}. This dataset comprises 8,976 sign language data samples with a resolution of 480*720, making it the largest dataset for recognizing long sentences in Chinese sign language to date.

The dataset's content focuses on international news, with sign language translations performed by professionally trained sign language interpreters, ensuring data accuracy. Notably, the dataset reflects real-world application scenarios by including instances where some nouns may have missing sign language counterparts, while their corresponding labels remain complete news sentences. Both the original video data and the landmark data processed by MediaPipe will be publicly released, contributing to the advancement of Chinese sign language recognition research.

\begin{table*}[ht]
	\centering
	\resizebox{\textwidth}{!}{
		\begin{tabular}{lccccccc}
			\toprule
			\textbf{Methods} & \multicolumn{2}{c}{\textbf{PHOENIX14}} & \multicolumn{2}{c}{\textbf{PHOENIX14T}} & \textbf{Mean CPU} & \textbf{FLOPs} & \textbf{Input Shape} \\
			& \textbf{Dev (\%)} & \textbf{Test (\%)} & \textbf{Dev (\%)} & \textbf{Test (\%)} & \textbf{(milliseconds)} & & \\
			\hline
			& \textbf{WER} & \textbf{WER} & \textbf{WER} & \textbf{WER} & & & \\
			\hline
			SFL \cite{35} & 26.2 & 26.8 & 25.1 & 26.1 & - & - & - \\
			FCN \cite{36} & 23.7 & 23.9 & 23.3 & 25.1 & - & - & - \\
			CMA \cite{37} & 21.3 & 21.9 & - & -& - & - & - \\
			VAC \cite{38} & 21.2 & 22.3 & - & -& - & $5 \times 10^{11}$ & - \\
			SMKD \cite{39} & 20.8 & 21 & 20.8 & 22.4 & - & - & - \\
			TLP \cite{40} & 19.7 & 20.8 & 19.4 & 21.2 & - & $4 \times 10^{11}$ & - \\
			SEN \cite{41} & 19.5 & 21 & 19.3 & 20.7 & - & - & - \\
			CNN+LSTM+HMM \cite{42} & 26 & 26 & 22.1 & 24.1 & - & - & - \\
			DNF* \cite{43} & 23.1 & 22.9 & - & -& - & $5 \times 10^{11}$ & - \\
			STMC* \cite{44} & 21.1 & 20.7 & 19.6 & 21.0 & - & $9 \times 10^{10}$ & - \\
			C2SLR* \cite{29} & 20.5 & 20.4 & 20.2 & 20.4 & - & $8 \times 10^{10}$ & - \\
			CorrNet \cite{12} & 18.8 & 19.4 & 18.9 & 20.5 & 5484.94 & $5 \times 10^{11}$ & [T,3,224,244] \\
			\hline
			&  & \multicolumn{2}{c}{\textbf{Landmark base}}  & & & &  \\
			\hline
			\textbf{MSLR} & 39.6 & 41.2 & 40.2 & 41.1 & \textbf{28.02} & $4 \times 10^{8}$ & [T,276] \\
			\textbf{KD-MSLR} & 28.7 & 30.7 & 29.6 & 31.2 & \textbf{28.02} & $4 \times 10^{8}$ & [T,276] \\
			\textbf{KD-MSLRT} & \textbf{17.4} & \textbf{17.8} & \textbf{16.9} & \textbf{18.4} & \textbf{42.64} & $\mathbf{1.9 \times 10^{9}}$ & [T,276] \\
			\bottomrule
	\end{tabular}}
	\caption{Comparison with the SOTA video-based SLR model on the PHOENIX 14 and PHOENIX14T datasets. * indicates that both video and landmark data types are used. Results better than the SOTAs are in bold.}
	\label{table1}
\end{table*}

\section{Experiments}

\subsection{Experiments Setup}

\subsubsection{Dataset}
To ensure a fair comparison, the model and data augmentation algorithms proposed in this research will be compared with SOTA methods on the PHOENIX14 \cite{koller2015continuous} and PHOENIX14T \cite{14} datasets. 

PHOENIX14 is recorded from a German weather forecast broadcast with nine actors before a clean back-ground with a resolution of 210 × 260. It contains 6841 sentences with a vocabulary of 1295 signs, divided into 5672 training samples, 540 development (Dev) samples and 629 testing (Test) samples. PHOENIX14-T is a dataset that comprises 8,247 sentences with a vocabulary of 1,085 signs, divided into 7,096 training instances, 519 development instances, and 642 testing instances.

\subsubsection{Training details}
We trained our model for 80 epochs with an initial learning rate of 0.0001, using a weight ratio of 1:25 for CTC loss and KD loss. We used the Adam optimizer with a weight decay of 0.0001 and employed a linear learning rate decay. For the input frames of the teacher network, we cropped them to 224*224 pixels and applied horizontal flipping on 50\%, rotation on 30\%, and temporal scaling on 20\% of the frames. For the landmarks of the student network, we applied horizontal flipping on 50\%, rotation on 30\%, and frame fusion on 20\% of the frames. Model training and evaluation were conducted on an Nvidia RTX 3080 graphics card, while CPU speed tests were performed on an Intel 13th generation i7 processor.

\subsubsection{Evaluation Metric}
In the CSLR field, the Word Error Rate (WER) serves as a common evaluation metric. It categorizes errors in predicted sentences into substitutions, insertions, and deletions, then calculates the minimum sum of these errors as the evaluation standard as:
\begin{equation}
	{WER} = \frac{\#{substitutions} + \#{insertions} + \#{deletions}}{\#{glosses\ \ in\ \ reference}},
\end{equation}
Note: The smaller the WER, the better the performance.

\subsection{Comparison with SOTA Methods:}
Table 1 gives the comparison between our KD-MSLRT model and other SOTA models. Currently, SOTA models on the PHOENIX14 and PHOENIX14T datasets are all based on video data format. Therefore, these models have larger parameters and computational requirements, resulting in relatively longer inference times. In contrast, our proposed model is based on landmark data extracted using MediaPipe, which compresses the data size by nearly 700 times. Our KD-MSLRT model reduces FLOPs by 260 times compared to SOTA models, significantly improving inference speed and runtime by 130 times. The TensorFlow Lite (TFLite) quantized model size is reduced to 12.93 MB, achieving millisecond-level inference speed on CPUs. Additionally, the KD-MSLRT model achieves a WER of 17.4\% (17.8\%), a 1.4\% (1.6\%) improvement on Dev (Test) set of PHOENIX14, and 16.9\% (18.4\%), a 2\% (2.1\%) improvement on PHOENIX14T. Although the FLOPs of KD-MSRLT increases significantly after applying the text correction model, the input of the text correction model consists of only 10-30 characters, resulting in minimal computational overhead, with slight increase of 14 milliseconds in the runtime. 

\subsection{Industrial Deployment}
We conducted a simulated deployment on mainstream development boards within the Intel i5 platform to validate the feasibility and cost-effectiveness of deploying our model in industrial environments. Table 2 presents the performance of our model on the PHOENIX14 and PHOENIX14T datasets at FP32 and INT8 precision levels, while Table 3 details the frames per second (FPS) achieved by our model during inference. The experimental results demonstrate that at INT8 precision, the model achieves a Word Error Rate (WER) comparable to that of FP32, with significantly lower computational costs. We further evaluated the model's inference speed, and the results indicate that on the Intel i5 platform, the model achieves high FPS across different precision levels, meeting the real-time requirements of industrial applications. The efficient inference at INT8 precision suggests that deployment on edge computing devices is feasible and can potentially reduce operational costs. This makes our model highly cost-effective for industrial deployment and suitable for large-scale application.

\begin{table}[h]
\setlength{\tabcolsep}{2.5pt}  
	\centering
	\resizebox{\columnwidth}{!}{
		\begin{tabular}{cccccc}
			\toprule
			\textbf{Datasets} & \textbf{FP32 (WER)} & \textbf{INT8 (WER)} & \textbf{Diff} & \textbf{Orisize} & \textbf{Onnxsize} \\ \hline
			PHOENIX14 & 17.4\%/17.8\% & 17.5\%/17.9\% & 0.5\%/0.4\% & 22.34MB & 12.93MB \\ 
			PHOENIX14T & 16.9\%/18.4\% & 16.8\%/18.5\% & -0.1\%/0.1\% & 23.84MB & 13.81MB \\ \hline
		\end{tabular}
	}
	\caption{Model performance comparison between FP32 and INT8 quantization.}
	\label{tab:model_performance}
\end{table}

\begin{table}[h]
\vspace{-2em}
	\centering
	\resizebox{\columnwidth}{!}{
		\begin{tabular}{cccc}
			\toprule
			\textbf{Datasets} & \textbf{FP32 (frame/s)} & \textbf{INT8 (frame/s)} & \textbf{Ratio (INT8/FP32)} \\ \hline
			PHOENIX14 & 11678 & 31369 & 2.69 \\ 
			PHOENIX14T & 10892 & 30154 & 2.74 \\ \hline
		\end{tabular}
	}
	\caption{FPS achieved by the model during inference.}
	\label{tab:single_core_performance}
\end{table}
\vspace{-1.5em}

\subsection{Ablation Experiments}
We present the results of ablation experiments on the development (Dev) and testing (Test) sets of the PHOENIX14 and PHOENIX14T.
\subsubsection{Exploring the effects of data augmentation, text correction (TC), and knowledge distillation (KD) on MSLR performance.}Table 4 gives that the WER of using only the text correction network is not as effective as when knowledge distillation and text correction network are used together. This is because the text correction model adjusts sentences based on the most dominant meaning expressed in the text. When the model's recognized text has a WER of over 40\%, the sentence may contain multiple conflicting meanings. If the number of words representing the incorrect meaning exceeds those representing the correct meaning, the text correction model may revise the text to reflect the incorrect meaning. Therefore, the text correction network performs better when sentences contain a sufficient amount of accurate information. 

\begin{table}[htb]
	\centering
	\resizebox{0.85\columnwidth}{!}{
		\begin{tabular}{cccccc}
			\toprule
			\textbf{Data Augmentation} & \textbf{KD} & \textbf{TC} & & \textbf{Dev (\%)} & \textbf{Test (\%)} \\
			\midrule
			\texttimes & \texttimes & \texttimes && 43.8/42.7 & 45.2/45.6 \\
			\checkmark & \texttimes & \texttimes && 39.6/38.2 & 41.2/40.7 \\
			\checkmark & \texttimes & \checkmark && 30.2/31.4 & 31.7/31.9 \\
			\texttimes & \checkmark & \texttimes && 28.7/28.4 & 30.7/29.3 \\
			\texttimes & \texttimes & \checkmark && 31.4/30.9 & 32.7/32.1 \\
			\texttimes & \checkmark & \checkmark && 18.2/17.9 & 18.7/19.1 \\
			\checkmark & \checkmark & \checkmark && \textbf{17.4/16.9} & \textbf{17.8/18.4} \\
			\hline
		\end{tabular}
	}
	\caption{Ablations for the performance improvement of MSLR by different modules. The best results are in bold.}
	\label{table5}
\end{table}

\subsubsection{Exploring the effects of single Transformer, dual Transformer, and data preprocessing on KD-MSLRT performance.} 
Table 5 gives the results of ablation experiments conducted on different modules of KD-MSLRT. The use of a dual Transformer (tf) significantly improves the model’s WER. Additionally, performing data preprocessing on the raw input text achieved further improvements of 0.9\% (1\%) and 1.2\% (0.7\%) on the Dev (Test) set of PHOENIX14 and PHOENIX14T, respectively.
\begin{table}[htb]
	\centering
	\resizebox{0.85\columnwidth}{!}{
		\begin{tabular}{ccccc}
			\toprule
			\textbf{Single tf} & \textbf{Dual tf} & \textbf{Pre-process} & \textbf{Dev (\%)} & \textbf{Test (\%)} \\
			\midrule
			\checkmark & \texttimes & \texttimes & 22.1/21.3 & 23.2/22.4 \\
			\checkmark & \checkmark & \texttimes & 18.3/18.1 & 18.8/19.1 \\
			\checkmark & \texttimes & \checkmark & 20.1/19.3 & 20.6/19.9 \\
			\checkmark & \checkmark & \checkmark & \textbf{17.4/16.9} & \textbf{17.8/18.4} \\
			\hline
		\end{tabular}
	}
	\caption{Ablations for the performance improvement of KD-MSLRT by different modules. The best results are in bold.}
	\label{table2}
\end{table}

\subsubsection{Research on the impact of 1D CNN convolutional kernel size on landmark data processing performance. }In Table 6, the results indicate that in the MSLR model, large convolutional kernels perform poorly when processing landmark data. The optimal kernel size is $K=5$, achieving WER of 43.8\% (45.2\%) and 42.6\% (44.8\%) on the Dev (Test) set of two datasets, respectively. Although using only the basic MSLR model yields significant advantages in terms of inference speed and parameter size, there is a considerable gap in performance compared to models based on video. This is because a significant amount of spatial information are lost during the process of converting video data to landmark.
\begin{table}[htb]
	\centering
	\resizebox{0.8\columnwidth}{!}{
		\begin{tabular}{lcc}
			\toprule
			\textbf{Configurations} & \textbf{Dev (\%)} & \textbf{Test (\%)} \\
			\hline
			$K=3$ & 45.7/44.9 & 46.8/47.1 \\
			$K=5$ & 43.8/42.6 & 45.2/44.8 \\
			$K=7$ & 47.9/46.3 & 48.7/47.3 \\
			$K=9$ & 49.1/48.2 & 50.9/50.3 \\
			\hline
		\end{tabular}
	}
	\caption{Ablations for the number of 1DCNN convolutional kernels on PHOENIX14 and PHOENIX14T datasets.}
	\label{table3}
\end{table}

\subsubsection{Exploring the effect of knowledge distillation on MSLR improvement.}Knowledge distillation improves model performance by transferring knowledge from a complex teacher model to a simpler student model, enhancing accuracy and efficiency. Using only convolutional layer knowledge distillation ($KD_{conv}$) in the Table 7 resulted in MSLR improving by 9.2\% (9.1\%) and 8.8\% (8.4\%) on Dev (Test) set of two datasets respectively. Using only knowledge distillation of BiLSTM ($KD_{bi}$) led to MSLR improving by 13.6\% (13.8\%) and 11.9\% (14.7\%) on Dev (Test) set. However, the improvement from using only self distillation ($KD_{self}$) was not significant, with only a 4.3\% (4\%) and 2.6\% (5.2\%) improvement on Dev (Test) set. It shows that using a 3D model as the teacher network for distilling knowledge into a 1D model significant improvements. The combined use of all three resulted in a 15.1\% (15.1\%) and 13.6\% (16.2\%) improvement on Dev (Test) set.
\begin{table}[htb]
	\centering
	\resizebox{0.85\columnwidth}{!}{
		\begin{tabular}{cccccc}
			\toprule
			\textbf{$KD_{conv}$} & \textbf{$KD_{bi}$} & \textbf{$KD_{self}$} && \textbf{Dev (\%)} & \textbf{Test (\%)} \\
			\midrule
			\texttimes & \texttimes & \texttimes && 43.8/42.7 & 45.2/45.6 \\
			\checkmark & \texttimes & \texttimes && 34.6/33.9 & 36.1/37.2 \\
			\checkmark & \texttimes & \checkmark && 32.1/31.4 & 34.3/34.9 \\
			\texttimes & \checkmark & \texttimes && 30.2/30.8 & 31.4/30.9 \\
			\texttimes & \texttimes & \checkmark && 39.5/40.1 & 41.2/40.4 \\
			\texttimes & \checkmark & \checkmark && 29.3/29.6 & 30.7/31.2 \\
			\checkmark & \checkmark & \checkmark && \textbf{28.7/29.1} & \textbf{30.1/29.4} \\
			\hline
		\end{tabular}
	}
	\caption{Performance of Base MSLR with different KD configurations. The best results are in bold.}
	\label{table4}
\end{table}

\section{Conclusion}
In this research, we propose a lightweight SLR model based on mediapipe and 3D to 1D knowledge distillation. It addresses the issues of high resource consumption, long runtime, and high deployment costs associated with video-based SLR models. Furthermore, to address the gap in augmentation methods for landmark data, we propose augmentation methods specifically tailored for landmark data based on techniques used for video and image augmentation. Additionally, to address the current lack of long-sentence datasets in the sign language pattern recognition field, we release a dataset containing 8,976 long-sentence sign language samples. We will make all aspects of this work publicly available, including datasets, in hope that these efforts will provide valuable insights for future research. 

\bibliography{aaai25.bib}

\end{document}